# NEW ADVANCES IN BAYESIAN CALCULATION FOR LINEAR AND NONLINEAR INVERSE PROBLEMS

A. Mohammad-Djafari, H. Carfantan and M. Nikolova
*Laboratoire des Signaux et Systèmes (CNRS-ESE-UPS)*
*École Supérieure d'Électricité,*
*Plateau de Moulon, 91192 Gif-sur-Yvette, France.*
*E-mail: djafari@lss.supelec.fr*

**Abstract.** The Bayesian approach has proved to be a coherent approach to handle ill posed Inverse problems. However, the Bayesian calculations need either an optimization or an integral calculation. The maximum a posteriori (MAP) estimation requires the minimization of a compound criterion which, in general, has two parts: a data fitting part and a prior part.

In many situations the criterion to be minimized becomes multimodal. The cost of the Simulated Annealing (SA) based techniques is in general huge for inverse problems. Recently a deterministic optimization technique, based on Graduated Non Convexity (GNC), have been proposed to overcome this difficulty.

The objective of this paper is to show two specific implementations of this technique for the following situations:
– Linear inverse problems where the solution is modeled as a piecewise continuous function. The non convexity of the criterion is then due to the special choice of the prior;
– A nonlinear inverse problem which arises in inverse scattering where the non convexity of the criterion is due to the likelihood part.

## 1. Introduction

We consider the case of general inverse problems:

$$\boldsymbol{y} = \boldsymbol{A}(\boldsymbol{x}) + \boldsymbol{b}, \tag{1}$$

where $\boldsymbol{x}$ is the vector of unknown variables, $\boldsymbol{y}$ is the data, $\boldsymbol{A}$ is a linear or non linear operator and $\boldsymbol{b}$ represents the errors which are assumed, hereafter, additive, zero-mean, white and Gaussian.

The Bayesian approach has proved to be a coherent approach to handle these problems. However, the Bayesian calculations need either an optimization or an integral calculation. The maximum a posteriori (MAP) estimation needs the minimization of a compound criterion:

$$J(\boldsymbol{x}) = -\log p(\boldsymbol{x}|\boldsymbol{y}) = Q(\boldsymbol{x}) + \lambda \Omega(\boldsymbol{x}) \tag{2}$$

which, in general, has two parts: a data fitting part:

$$Q(\boldsymbol{x}) = -\log p(\boldsymbol{y}|\boldsymbol{x}) = \|\boldsymbol{y} - \boldsymbol{A}(\boldsymbol{x})\|^2 \tag{3}$$

and a prior part:

$$\Omega(\boldsymbol{x}) = -\log p(\boldsymbol{x}) = \sum_r \sum_{s \in \mathcal{N}_r} \phi(x_r - x_s). \qquad (4)$$

This last expression is due to general Markov modeling where $r$ and $s$ are two site indexes and $\mathcal{N}_r$ means neighbor sites of $r$.

In many situations the criterion $J(\boldsymbol{x})$ becomes multimodal. We consider here two cases: The first is the case of general linear inverse problems with markovian priors with non convex energies, and the second is, the case of non-linear inverse problem.

In both cases we need a global optimization technique to determine the solution. In the first case the non convexity is due to the second termed and in the second case the non convexity is more due to the first termed.

The cost of the Simulated Annealing (SA) based techniques is mainly dependent to the neighborhood size of the posterior marginal probability distribution $p(x_j|\boldsymbol{x},\boldsymbol{y})$ which is directly related to the neighborhood size of the prior marginal probability distribution $p(x_j|\boldsymbol{x})$ and the support of the operator $\boldsymbol{A}$. When $\boldsymbol{A}$ is a local operator with very small support, i.e.; when the data element $y_i$ depends only to a few number of the unknown variables $x_j$, then SA can be implemented efficiently [1, 2, 3]. But, unfortunately, this is not the case of many inverse problems, where the support of the operator is not small. The cost of SA is then, in general, huge for these problems.

Recently a deterministic relaxation algorithm, inspired by the Graduated Non Convexity (GNC) principle, has been proposed by Blake and Zisserman in [4, 5] for the optimization of the multimodal MAP criteria. They have shown its efficiency in practical applications for noise cancellation and segmentation. This algorithm has been extended to the general linear ill-posed inverse problem by Nikolva *et al.* in [6].

The object of this presentation is to show two specific implementations of this technique for two specific cases of the two aforementioned situations, i.e.;
– The linear inverse problems where the solution is modeled as a piecewise continuous function using a compound markov modeling (for example the intensity and the line process in image reconstruction), where the non convexity of the criterion is due to the markovian priors; and
– A special non-linear inverse problem which arises in inverse scattering and diffraction tomography imaging applications where the non convexity of the criterion is due to the likelihood part.

The paper is organized as follows: The next section presents the main idea of the GNC principle. Sections 3 and 4 will consider the two aforementioned specific cases and, finally, some simulation results will illustrate the performances of the proposed method in two special applications.

## 2. Graduated Non Convexity scheme

The principle of this algorithm is very simple. It consists of approximating the non convex criterion $J(\boldsymbol{x})$ with a sequence of continuously derivable criteria $J_{c_k}(\boldsymbol{x})$ such that:
- the first one $J_{c_0}(\boldsymbol{x})$ be convex;
- the final one (the limit) $J_{c_k}(\boldsymbol{x})$ converges to $J(\boldsymbol{x})$:

$$\lim_{k \mapsto \infty} J_{c_k}(\boldsymbol{x}) = J(\boldsymbol{x}), \quad \forall \boldsymbol{x},$$

where $c_k > 0$ are increasing relaxation parameters; and then
- for each $k$, a relaxed solution is calculated by minimizing locally, initialized by the previous solution, as follows:

$$\widehat{\boldsymbol{x}}_{c_k} = \arg \min_{\boldsymbol{x} \in V(\widehat{\boldsymbol{x}}_{c_{k-1}})} \{J_{c_k}(\boldsymbol{x})\} \tag{5}$$

Fig. 2. illustrates such a scheme. The hope is that the sequence $\widehat{\boldsymbol{x}}_{c_k}$ converges to the global minimizer of $J(\boldsymbol{x})$. Note that there is no theoretical ground for this hope, however, in many practical applications it seems to be realist.

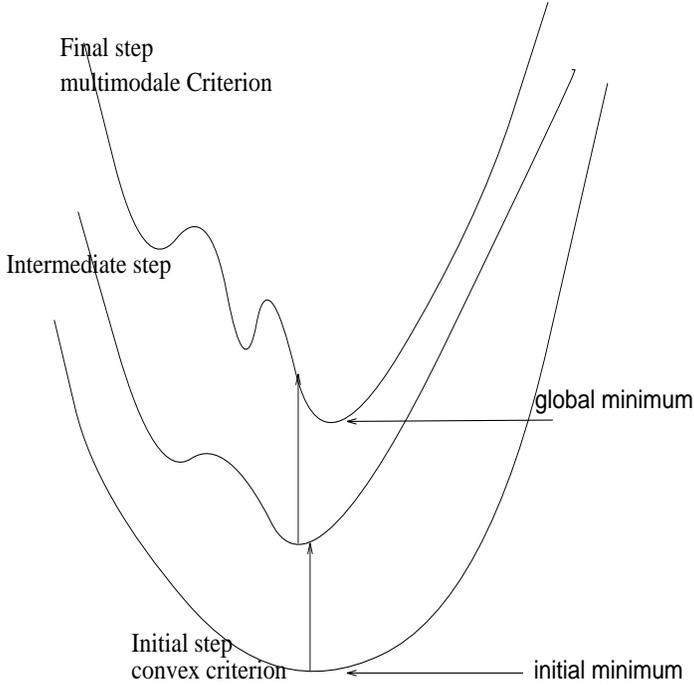

Figure 1: GNC scheme.

### 3. Linear inverse problems with a piecewise Gaussian prior

Let first consider a very simple noise filtering problem:

$$\boldsymbol{y} = \boldsymbol{x} + \boldsymbol{b}, \tag{6}$$

where we know that $\boldsymbol{x}$ represents the samples of a piecewise continuous function $x(t)$. Blake and Zisserman in [4] proposed to estimate $\boldsymbol{x}$ by searching the global minimum of the following criterion:

$$J(\boldsymbol{x}) = \|\boldsymbol{y} - \boldsymbol{x}\|^2 + \Omega(\boldsymbol{x}) \tag{7}$$

with

$$\Omega(\boldsymbol{x}) = \sum_j \phi(t_j), \quad t_j = x_j - x_{j-1} \tag{8}$$

and

$$\phi(t) = \begin{cases} (\lambda t)^2 & \text{if } |t| < T \\ \alpha = (\lambda T)^2 & \text{if } |t| > T \end{cases} \tag{9}$$

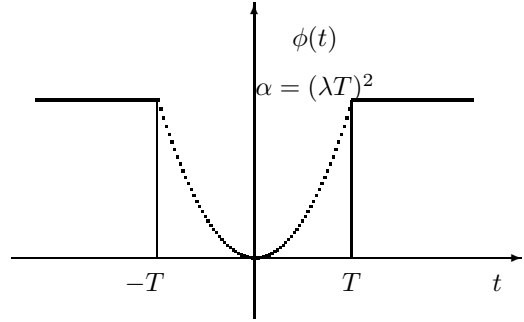

Figure 2: Truncated quadratic function.

Note that, for $\phi(t) = t^2$ the criterion $J(\boldsymbol{x})$ can be considered either as the MAP criterion with Gaussian prior or as the Tikhonov regularization one. The choice (9) for $\phi(t)$ is done to preserve the discontinuities in $x(t)$. With this choice, obviously, $J(\boldsymbol{x})$ is multimodal. The GNC idea was then to construct:

$$J_c(\boldsymbol{x}) = \|\boldsymbol{y} - \boldsymbol{x}\|^2 + \Omega_c(\boldsymbol{x}) \tag{10}$$

with

$$\Omega_c(\boldsymbol{x}) = \sum_j \phi_c(t_j), \tag{11}$$

and

$$\phi_c(t) = \begin{cases} (\lambda t)^2 & \text{if } |t| < q_c \\ \alpha - \tfrac{1}{2}c(|t| - r_c)^2 & \text{if } q_c > |t| > r_c \\ \alpha = (\lambda T)^2 & \text{if } |t| > r_c \end{cases}, \quad \begin{cases} q_c = T(1 + 2\lambda^2/c)^{-1/2} \\ r_c = T(1 + 2\lambda^2/c)^{1/2} \end{cases} \tag{12}$$

and find a $c_0$ such that $J_c$ for $c < c_0$ be convex and then for a given sequence of relaxations parameters $\{c_0, c_1, \ldots, c_k\}$ do:

$$\widehat{\boldsymbol{x}}_{c_k} = \arg \min_{\boldsymbol{x} \in V(\widehat{\boldsymbol{x}}_{c_{k-1}})} \{J_{c_k}(\boldsymbol{x})\} \tag{13}$$

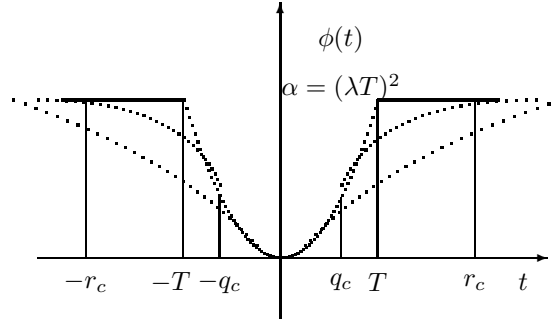

Figure 3: Truncated quadratic function and it's relaxation scheme.

Blake and Zisserman in [4] showed the existence of $c_0$ such that $J_{c_0}(\boldsymbol{x})$ be convex and global convergence of this algorithm. However, for the case of inverse problems $\boldsymbol{y} = \boldsymbol{A}\boldsymbol{x} + \boldsymbol{b}$ where $\boldsymbol{A}$ is singular or ill-conditioned, the existence of $c_0$ is no more insured. Nikolova *et al.* [6, 7] extended this work by proposing a doubly relaxed criterion:

$$J_c(\boldsymbol{x}) = \|\boldsymbol{y} - \boldsymbol{x}\|^2 + \Omega_{a,c}(\boldsymbol{x}) \tag{14}$$

with

$$\Omega_{a,c}(\boldsymbol{x}) = \sum_j \left[\phi_c(t_j) + at_j^2\right] \tag{15}$$

and the following double relaxation scheme:

$$\text{for fixed } c = c_0 \quad \text{and for } a = a_0, \ldots, 0 \text{ do:}$$
$$\widehat{\boldsymbol{x}}_{a_k} = \arg \min_{\boldsymbol{x} \in V(\widehat{\boldsymbol{x}}_{a_{k-1}})} \{J_{a_k,c_0}(\boldsymbol{x})\} \tag{16}$$

$$\text{for fixed } a = 0 \quad \text{and for } c = c_0, \ldots, \infty \text{ do:}$$
$$\widehat{\boldsymbol{x}}_{c_k} = \arg \min_{\boldsymbol{x} \in V(\widehat{\boldsymbol{x}}_{c_{k-1}})} \{J_{c_k,0}(\boldsymbol{x})\} \tag{17}$$

The initial convexity of the criterion is insured for $a_0 > \frac{c_0}{2}$. Many details, discussions and more extensions, specially for 2D case, are given in [7, 8, 9].

3.1. Link with compound Markov models

Compound Markov modeling in image processing became popular after the works of Geman & Geman [1] and Besag [10, 11]. To see the link with these model briefly, let consider a 1-D case. When $x(t)$ is assumed piecewise continuous or piecewise Gaussian, it can be modeled with a set of coupled variables $(\boldsymbol{x}, \boldsymbol{l})$, where

$l$ is a vector of binary-valued variables and $x$ is Gaussian. Now, consider the MAP estimate of $(x, l)$:

$$(\widehat{x}, \widehat{l}) = \arg \min_{(x,l)} \{J(x, l) = -\log p(x, l|y)\} \tag{18}$$

where

$$-\log p(x, l|y) = -\log p(y|x) - \log p(x|l) - \log p(l) + cte, \tag{19}$$

with the following prior laws:

$$\begin{aligned} -\log p(y|x) &= \|y - Ax\|^2 \\ -\log p(x|l) &= \sum_j \phi(t_j)(1 - l_j), \quad t_j = x_j - x_{j-1}, \quad \phi(t) = (\lambda t)^2 \\ -\log p(l) &= \alpha \sum_j l_j \end{aligned}$$

we obtain:

$$(\widehat{x}, \widehat{l}) = \arg \min_{(x,l)} \{J(x, l)\} \tag{20}$$

with

$$J(x, l) = \|y - Ax\|^2 + \sum_j \left[\lambda^2 (x_j - x_{j-1})^2 (1 - l_j) + \alpha l_j\right]. \tag{21}$$

Note that the line variables $l_j$ are assumed non-interacting (mutually independent). With this hypothesis, it is easy to show that the solution $(\widehat{x}, \widehat{l})$ obtained by (20) is equivalent to the solution obtained by

$$\widehat{x} = \arg \min_x \{J(x)\} \quad \text{with} \quad J(x) = \|y - Ax\|^2 + \sum_j \phi(x_j - x_{j-1})^2 \tag{22}$$

with $\phi(t)$ defined by (9) and

$$\widehat{l}_j = \begin{cases} 1 & \text{if } |\widehat{x}_j - \widehat{x}_{j-1}| > T \\ 0 & \text{if } |\widehat{x}_j - \widehat{x}_{j-1}| < T \end{cases} \tag{23}$$

### 4. A non linear inverse problem

To show how the GNC principle can be used for nonlinear inverse problems we consider the case of inverse scattering and more specifically the diffraction tomography. To be short in presentation of the application, we give here an abstract presentation of the problem. For more details on derivation of the inverse scattering and diffraction tomography application see [12, 13, 14]. To summarize, considering the geometry of Fig. 4., we have the following relations:

$$y(r_i) = \iint_D G_m(r_i, r') \phi(r') x(r') \, dr', \quad r_i \in S, \tag{24}$$

$$\phi(r) = \phi_0(r) + \iint_D G_o(r, r') \phi(r') x(r') \, dr', \quad r \in D. \tag{25}$$

The discretized version of these two equations can be written with the following compact notations:

$$\boldsymbol{y} = \boldsymbol{G}_m \boldsymbol{X} \boldsymbol{\phi} + \boldsymbol{b}, \qquad (26)$$
$$\boldsymbol{\phi} = \boldsymbol{\phi}_0 + \boldsymbol{G}_o \boldsymbol{X} \boldsymbol{\phi}, \qquad (27)$$

where $\boldsymbol{\phi}_0$ is the incident field, $\boldsymbol{G}_m, \boldsymbol{G}_o$ are matrices related to the Green functions, $\boldsymbol{X}$ is a diagonal matrix with the components of the vector $\boldsymbol{x}$ (a $n$ length vector) as its diagonal elements and $\boldsymbol{y}, \boldsymbol{\phi}$ are respectively $m$ and $n$ length vectors representing the measured data (scattered field) and the total field on the object. Note that $n$ may be greater than $m$.

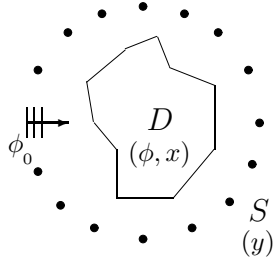

Figure 4: Diffraction tomography geometry configuration

These two equations can be combined to obtain a symbolic explicit relation between the data $\boldsymbol{y}$ and the unknowns $\boldsymbol{x}$:

$$\boldsymbol{y} = \boldsymbol{G}_m \boldsymbol{X} \left(\boldsymbol{I} - \boldsymbol{G}_o \boldsymbol{X}\right)^{-1} \boldsymbol{\phi}_0 + \boldsymbol{b} = \boldsymbol{A}(\boldsymbol{x}) + \boldsymbol{b}, \qquad (28)$$

where the considered matrix is assumed to be invertible. Now, the inverse problem we are faced is to find $\boldsymbol{x}$ given $\boldsymbol{y}$. Note that, $\boldsymbol{\phi}_0$, $\boldsymbol{G}_o$ and $\boldsymbol{G}_m$ are known and the relation between $\boldsymbol{x}$ and $\boldsymbol{y}$ is non linear. In fact, given $\boldsymbol{\phi}$, $\boldsymbol{y}$ is linear in $\boldsymbol{x}$ (26), but $\boldsymbol{\phi}$ depends on $\boldsymbol{x}$ through the second equation (27).

Here also, using the Bayesian approach, the MAP estimate is defined as the minimizer of

$$J(\boldsymbol{x}) = \|\boldsymbol{y} - \boldsymbol{A}(\boldsymbol{x})\|^2 + \Omega(\boldsymbol{x}) \qquad (29)$$

and, even when $\Omega(\boldsymbol{x})$ is chosen to be convex, $J(\boldsymbol{x})$ may not due to the fact that $\|\boldsymbol{y} - \boldsymbol{A}(\boldsymbol{x})\|^2$ is no more quadratic in $\boldsymbol{x}$. Carfantan *et al.* in [12, 13, 14] proposed to use the GNC idea in this case.

To introduce the GNC technique, they considered the following relaxation sequence:

$$\boldsymbol{A}_{c_k}(\boldsymbol{x}) = \boldsymbol{G}_m \boldsymbol{X} (\boldsymbol{I} - c_k \boldsymbol{G}_o \boldsymbol{X})^{-1} \boldsymbol{\phi}_0, \qquad (30)$$

with $c_0 = 0$, and $\lim_{k \to \infty} c_k = 1$.

Note that the first term ($c_0 = 0$) corresponds to a linearized model for the problem named the Born approximation which consists in neglecting partially the diffraction effects. This results to the following convex criterion:

$$J_0(\boldsymbol{x}) = \|\boldsymbol{y} - \boldsymbol{G}_m \boldsymbol{X} \boldsymbol{\phi}_0\|^2 + \Omega(\boldsymbol{x}).$$

Note also that for $c_k = 1$ the criterion $J_1(\boldsymbol{x}) = J(\boldsymbol{x})$. The main practical problem is then the choice of sequences $\{c_k = 0, \cdots, 1\}$ which is done by experiment.

## 5. Some simulation results and applications

### 5.1. 1-D NOISE FILTERING

Fig. 5.1. shows an example of results obtained in noise filtering. In this figure we see the original signal, noisy data, restoration using a Gaussian model ($\phi(t)$ quadratic) and restoration obtained by GNC when $\phi(t)$ is chosen to be truncated quadratic.

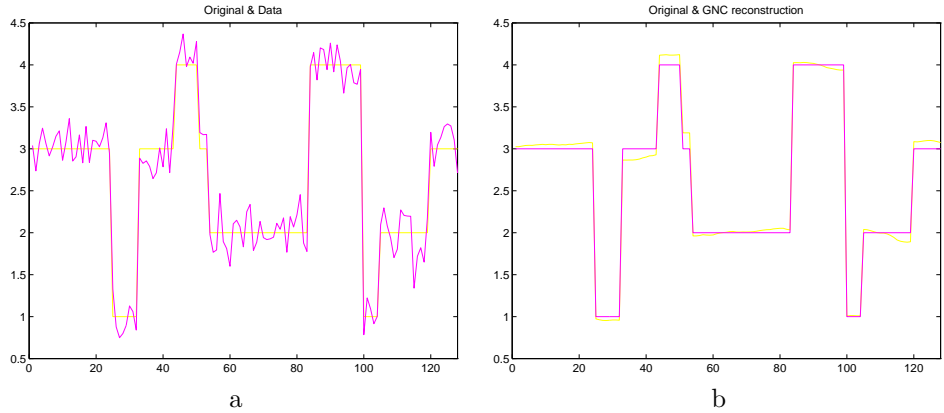

Figure 5: Noise filtering:
a) Original and data, b) Original, Gaussian restoration and GNC restoration

### 5.2. 1-D SIGNAL DECONVOLUTION

Fig. 5.2. shows an example of results in signal deconvolution. In this figure we see the original signal, noisy data, restoration using a Gaussian model and restoration obtained by GNC.

### 5.3. IMAGE RESTORATION

Fig. 5.3. shows an example of results in Image restoration. In this figure a) is the original image, b) is the blurred and noisy data, c) is the restoration using a Gaussian model and d) is the restoration obtained by GNC with truncated quadratic regularization.

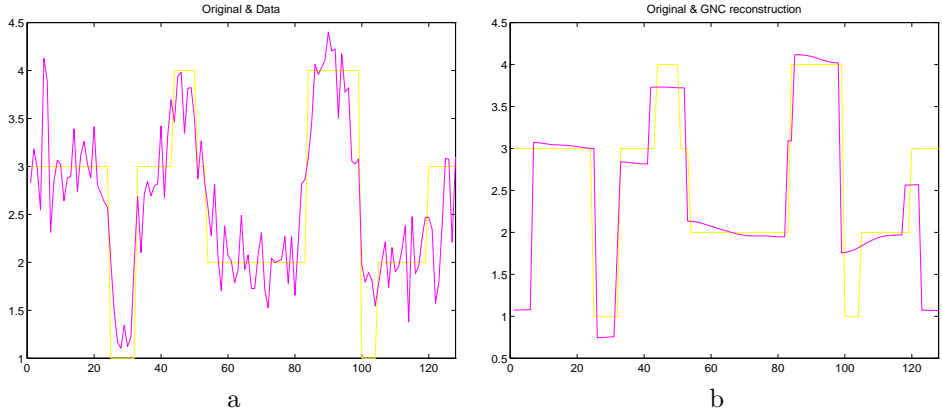

Figure 6: Deconvolution:
a) Original and data, b) Original, Gaussian restoration and GNC restoration

5.4. IMAGE RECONSTRUCTION IN X-RAY TOMOGRAPHY

Fig. 5.4. shows results obtained in X-ray tomography image reconstruction. In this figure a) shows the original image, b) shows the projections (data), c) shows the backprojection reconstruction, d) shows a reconstruction using a Gaussian prior, e) shows a reconstruction using a Gamma prior, and f) shows a reconstruction using GNC with truncated quadratic regularization.

5.5. INVERSE SCATTERING AND DIFFRACTION TOMOGRAPHY

Fig. 5.5. shows an example of results in non linear diffraction tomography image reconstruction. In this figure a) is the original image, b) is the measured scattered field data, d) is a reconstruction using the linear Born approximation, and e) is a reconstruction using GNC.

6. Conclusions

The Bayesian maximum *a posteriori* (MAP) estimates requires the minimization of a compound criterion which, in general, has two parts: a data fitting part and a prior part. In many situations in inverse problems the criterion to be minimized is multimodal. The cost of the Simulated Annealing (SA) based techniques is in general huge for these problems.

We reported here recently proposed new techniques, based on Graduated Non Convexity (GNC), to overcome this difficulty and showed two specific implementations of this technique for:

– The linear inverse problems such as: noise filtering, deconvolution, image restoration and tomographic image reconstruction, where the solution is modeled as a piecewise continuous function and where, the non convexity of the criterion is due to the special choice of the prior; and

– A nonlinear inverse inverse scattering and diffraction tomography where the non convexity of the criterion is due to the likelihood part.

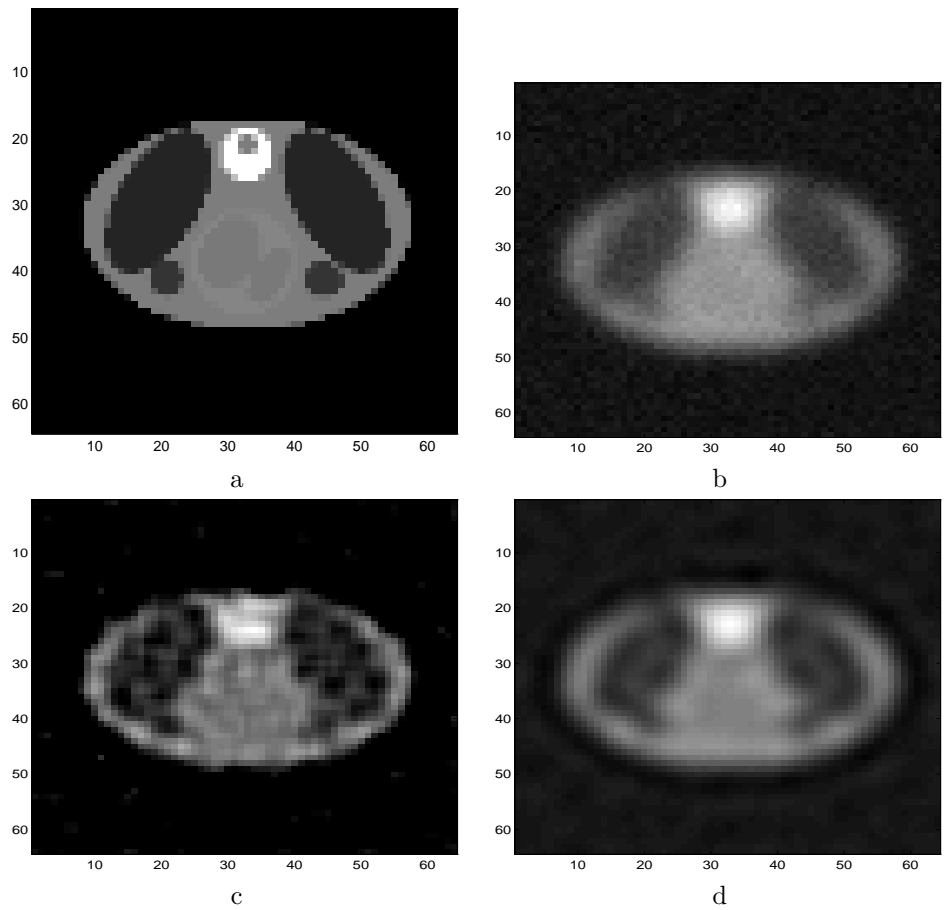

Figure 7: Image restoration:
a) original, b) data, c) Gaussian restoration, d) GNC restoration

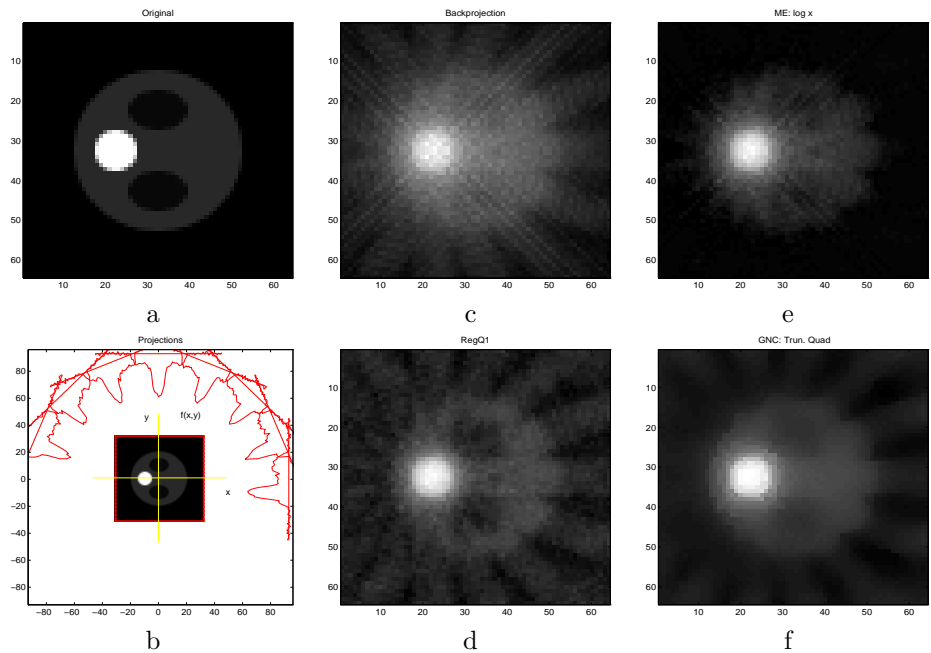

Figure 8: Image reconstruction in X-ray tomography:
a) original, b) projections (data), c) Backprojection, d) Gaussian reconstruction,
d) Gamma prior reconstruction, and e) GNC reconstruction

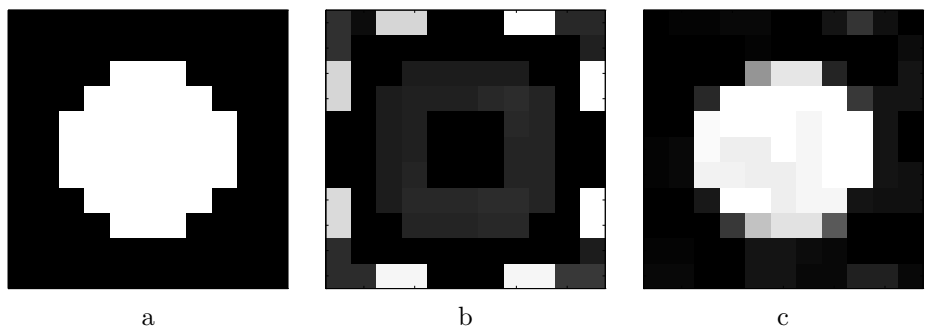

Figure 9: Diffraction tomography image reconstruction:
a) original, b) linear Born approximation reconstruction, c) GNC based reconstruction